# A Comprehensive Computational Framework for Materials Design, Ab Initio Modeling, and Molecular Docking

Md Rakibul Karim Akanda, Michael P. Richard

College of Engineering Technology and Computing, Savannah State University—Savannah, GA 31404, United States of America

**Abstract**

To facilitate rational molecular and materials design, this research proposes an integrated computational framework that combines stochastic simulation, ab initio quantum chemistry, and molecular docking. The suggested workflow allows systematic investigation of structural stability, binding affinity, and electronic properties across biological and materials science domains by utilizing complementary tools like Avogadro for molecular construction and visualization, AutoDock for docking and interaction analysis, and ORCA for high-level electronic structure computations. Uncertainty, configurational sampling, and optimization in high-dimensional chemical spaces are addressed by combining Monte Carlo-based and annealing-inspired techniques. The work shows how materials science ideas such as polymer design, thin films, crystalline lattices, and bioelectronic systems can be applied to drug development. On-device, open-source computational methods are viable, scalable, and economical, as demonstrated by comparative platform analysis. All things considered, the findings highlight the need of an integrated, repeatable computational pipeline for speeding up de novo molecule assembly and materials architecture while lowering experimental risk and expense.

1. Introduction

With only little variations in graphics API performance, Avogadro is a perfect interactive tool on both systems due to its low energy and disk space consumption [1]. Although users must deal with variations in installation and data handling techniques, AutoDock operates consistently in terms of energy and runtime on both platforms [2]. Of the three, ORCA is the most computationally demanding (Figure 1). Its performance is mostly determined by hardware rather than the operating system, and its energy consumption and disk space needs scale with the complexity of the calculation.

To improve collaboration and reproducibility, unified research initiatives like the Human Microbiome Project seek to provide a common framework for data exchange among various research organizations. An essential component of these ecosystems is Avogadro, an open-source molecular visualization tool that facilitates smooth integration with standardized data formats. Avogadro can read a variety of molecular data formats thanks to Open Babel integration, guaranteeing platform compatibility. Avogadro is a perfect tool for academics working on large-scale, collaborative projects because of its capacity to harmonize data, which enables them to see intricate molecular structures and carry out exploratory studies in a shared, repeatable pipeline (Figure 2). Additionally, because it is open-source, security measures like strong encryption and local downloads are upheld, which is crucial for safeguarding private study data.

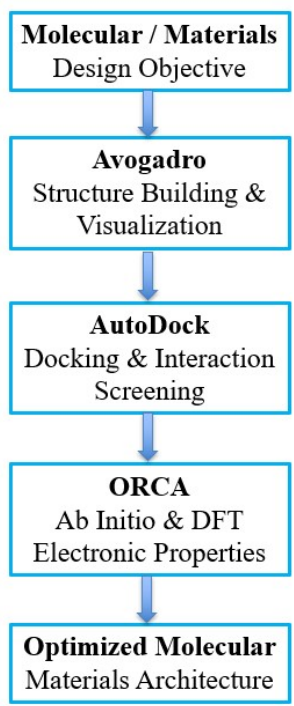

Figure 1. Integrated Computational Workflow

Avogadro also facilitates the creation of plugins designed for research projects, including the automatic retrieval of metabolomic or crystallographic data from central repositories. These plugins guarantee that researchers have access to the most recent data while reducing web traffic and upholding security by utilizing secure APIs with SSH encryption. Researchers from several labs or institutions can collaborate with the same dataset in multidisciplinary settings, comparing structural variations, visualizing molecular interactions, and performing force field optimizations. Avogadro is a vital tool for varied research teams trying to tackle challenging scientific problems since it integrates with computational tools like Jupyter Notebooks, Google Colab, and Domino to enable collaborative, real-time analysis.

Avogadro's ability to improve education and bridge disciplines extends beyond individual studies. The capacity to query and visualize chemical and biological data from structured libraries is essential in interdisciplinary research, such as integrating chemistry with microbiome studies or material science. Avogadro can enable chemists, biologists, and data scientists to view molecular data through a single lens by integrating data from databases including PubChem, HMDB, PDB, and UniProt. This capacity also extends to the classroom, where students can learn how to visualize complex molecules and comprehend how chemistry, biology, and computational science interact.

Avogadro guarantees that researchers and educators can preserve efficiency and repeatability by consolidating the data visualization and analysis process into a single platform.

The foundation of molecular docking is the idea of molecular recognition, in which the biological activity of two molecules is largely determined by their binding contacts (Figure 3). Important elements including hydrogen bonds, hydrophobic forces, and electrostatic interactions control this process. The search algorithm and the scoring function are the two main steps in the molecular docking process. Using techniques like genetic algorithms and molecular dynamics simulations, the search algorithm investigates potential ligand orientations and conformations within the receptor's binding region. The scoring procedure assesses each interaction's strength once possible binding postures are produced, yielding a numerical representation of binding affinity. Predicting which molecular conformations will bind to the target site most successfully is the main goal, which will direct the development of therapeutic medicines.

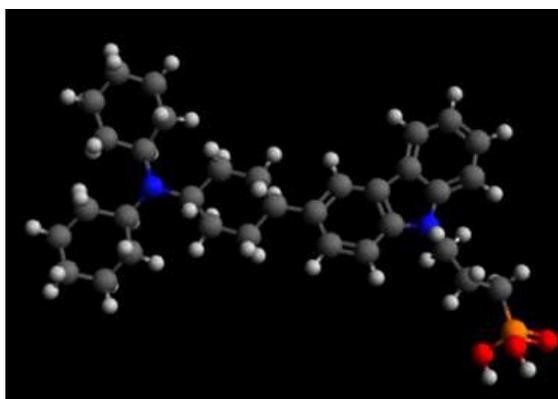

Figure 2. Avogadro Monolayer Film Compound

Depending on the data at hand and the objectives of the study, several molecular docking techniques provide unique benefits. Structure-based docking places the ligand into the binding site using the receptor's three-dimensional structure, which is discovered using experimental methods like X-ray crystallography or NMR. For systems with well-defined receptor structures, this strategy works quite well. However, when the structure of the receptor is unknown, ligand-based docking uses pharmacophore models and concentrates on known ligand structures. When receptor data is lacking or insufficient, this method is useful. By using both receptor architecture and ligand information to produce more precise binding predictions, hybrid docking combines the advantages of both approaches.

Molecular docking is used in many different domains, most notably medication research and discovery. By forecasting their binding affinity to target proteins, virtual screening makes it possible to quickly identify possible drug candidates, saving time and money in the early phases of drug development. Lead optimization uses docking to improve the structure of potential drugs, increasing their selectivity and binding efficiency. These techniques can also be used in material science to model molecular interactions in new materials, enhancing their suitability for uses. Furthermore, docking is essential for researching mechanisms of action and forecasting drug resistance, providing information that directs the creation of medications with improved efficacy and decreased resistance susceptibility.

A key component of computational chemistry, especially for the study of molecular electronic structures, is the extremely flexible quantum chemistry software program ORCA (Figure 4). ORCA, which was created by Frank Neese and his research team, offers a wide range of computational approaches, making it a priceless resource for scientists working in various fields [3]. It is particularly well known for its dependability and effectiveness in simulating the electrical structure of molecules, allowing everything from simple structure improvements to intricate, advanced research. Although ORCA is excellent at modeling systems of different levels of complexity, it also helps researchers with big or multi-atom systems, where computations can become costly.

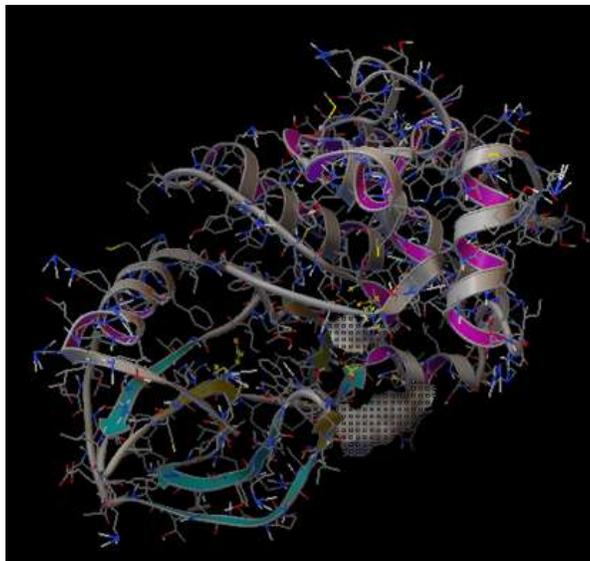

Figure 3. Auto Dock software capabilities of showing graphing complex features of molecules

Density Functional Theory (DFT), a popular technique in computational chemistry for examining the electronic structure of atoms, molecules, and solids, is one of the main characteristics that sets ORCA apart. By employing functionals of the electron density to approximate the many-body Schrödinger equation, DFT allows precise predictions of features including vibrational frequencies, electronic structure, and magnetic behavior. When it comes to solving computational problems like excited-state computations and lattice structure analysis, ORCA's DFT implementation excels. It can add dispersion corrections (e.g., DFT-D3) to more accurately describe noncovalent interactions and supports a wide range of functionals, including Generalized Gradient Approximation (GGA), hybrid, and meta-GGA functionals.

To handle the complexity of large-scale systems like lattices and periodic structures, ORCA's computing capabilities go beyond simple molecular systems. In these situations, ORCA calculates electronic band structures and examines material features such as dislocations and vibrations by using Bloch wave equations to describe the behavior of electrons in a periodic potential. Researchers can analyze periodicity and accurately estimate the electronic characteristics of materials by sampling reciprocal space using K-points in the Brillouin zone. By using frozen-core potentials and pseudopotentials, ORCA dramatically lowers computing costs for systems with periodic boundary conditions without sacrificing computation accuracy.

By offering an effective and economical means of predicting molecular and material properties, ORCA offers substantial advantages over conventional experimental approaches in pharmaceutical R&D, materials science, and academic research. By mimicking molecular characteristics like NMR chemical shifts and UV/Vi's absorption bands, ORCA enables researchers to evaluate and optimize compounds in medication development, eliminating the need for costly and time-consuming tests. For academic institutions, where ORCA is free, and pharmaceutical businesses, where cloud-based High-Performance Computing (HPC) clusters offer an affordable substitute for lab equipment maintenance, this computational technique is very beneficial. With the capacity to replicate intricate electronic transitions and lattice behavior, ORCA assists researchers in materials science by forecasting optical absorption spectra and other characteristics of organic semiconductors. This offers a quicker and more adaptable substitute for conventional spectrometric instruments, which can be expensive and time-consuming. ORCA is an invaluable instructional tool for academic research and teaching because of its speed and accessibility, which enable teachers and students to run simulations and compare them with experimental results from campus-based NMR equipment. While spectrometric instruments have drawbacks including availability, calibration, and sample preparation, ORCA makes simulations quick, repeatable, and scalable, making computational chemistry a vital tool for research and teaching.

```
--------------
DIPOLE MOMENT
--------------
                              X              Y              Z
Electronic contribution:   -0.09611        0.00737       -0.00000
Nuclear contribution    :   0.00713       -0.00055        0.00000
                         -----------------------------------------
Total Dipole Moment     :  -0.08898        0.00682       -0.00000
                         -----------------------------------------
Magnitude (a.u.)        :   0.08924
Magnitude (Debye)       :   0.22683

--------------------
Rotational spectrum
--------------------

Rotational constants in cm-1:     0.000000      1.930781      1.930781
Rotational constants in MHz :     0.000000  57883.356953  57883.356953

 Dipole components along the rotational axes:
x,y,z [a.u.] :     0.089242     -0.000000     -0.000000
x,y,z [Debye]:     0.226834     -0.000000     -0.000000

Timings for individual modules:

Sum of individual times         ...        2.321 sec (=   0.039 min)
GTO integral calculation        ...        0.246 sec (=   0.004 min)  10.6 %
SCF iterations                  ...        2.075 sec (=   0.035 min)  89.4 %
                    ****ORCA TERMINATED NORMALLY****
```

Figure 4. Orca test results

2. **Procedure**

Research on a variety of materials and technologies that help produce electronics that can be helpful in a variety of applications has been conducted during the past few decades [4–17]. The outcomes of each software after a few iterations may be shown below. A setup wizard installer package or a manual compilation method came with every piece of software. A license for one use and agreement was given for each piece of software, although the Orca package did require a user's forum account. Since only one of each type of software was used, there was no speed test.

Testing allows us to advance theory to the next level of application and potential, making de novo assembly feasible. When the software is tested on hardware that is comparable to every position in a corporate setting, it reveals that its user-friendly graphical interfaces make it reliable for both unskilled and specific users. Together, AutoDock, ORCA, and Avogadro simplify molecular development by offering a comprehensive computational workflow from preliminary design to in-depth analysis. Avogadro is the perfect place to start when creating models because it makes it easy to design and visualize molecular architectures. The interactions between these compounds and any biological or material targets are then predicted and optimized by AutoDock using advanced docking algorithms, which is essential for guaranteeing functional activity. In addition to these resources, ORCA provides comprehensive quantum chemical computations that improve our comprehension of electronic structure and energetic characteristics and enable accurate modification of molecular behavior. By minimizing experimental guessing and enabling quick iteration toward molecules with specific, crucial features, these software programs collectively improve development efficiency.

### 3. Conclusion

This study shows how combining AutoDock, ORCA, and Avogadro can significantly improve the logical assembly of molecules with specific key features. Avogadro offered an easy-to-use platform for molecular construction and excellent visualization, AutoDock facilitated the effective prediction of ligand-receptor interactions through sophisticated docking simulations, and ORCA supplied comprehensive quantum chemical insights that directed the fine-tuning of electronic properties. Each software contributed a distinct and crucial capability. By combining these complimentary approaches, we created a reliable computational pipeline that improves the accuracy of property optimization while also speeding up molecular design. We were able to close the gap between theoretical models and real-world applications by methodically identifying and improving viable molecular candidates thanks to this comprehensive methodology. This paradigm highlights how molecular modeling, docking, and quantum computations can be combined to spur innovation in fields including bioelectronics, materials science, and drug development. To further simplify the assembly process and handle increasingly complex molecular systems, future development will concentrate on boosting the pipeline's scalability and automation.

### Acknowledgements

This work was supported as part of the Modeling and Simulation Program (MSP) grant funded by the US Department of Education under Award No. P116S210002 and Improving Access to Cyber Security Education for Underrepresented Minorities funded by the US Department of Education under Award No. P116Z230007.